\documentstyle[twocolumn]{jpsj} 
\input epsf
\input{epsf.sty}

\def\runtitle{Vortices in C$_{60}$--irradiated BSCCO}
\def\runauthor{N. Ishikawa, C.J. van der Beek, A. Dunlop et al.}

\title{Vortex phase diagram in Bi$_{2}$Sr$_{2}$CaCu$_{2}$O$_{8+\delta}$ 
with damage tracks created by 30 MeV fullerene irradiation}
\author{\textsc{N. Ishikawa}$^{1,2}$, \textsc{C.J. van der Beek}$^{1}$
\footnote{Email address : kees.vanderbeek@polytechnique.fr}, 
\textsc{A. Dunlop}$^{1}$, \textsc{G. Jaskierowicz}$^{1}$,
\textsc{Ming Li}$^{3}$\footnote{present address : Philips Semiconductors, 
Gerstweg 2, 6534 AE Nijmegen, the Netherlands},
\textsc{P.H. Kes}$^{3}$, \textsc{S. Della-Negra}$^{4}$}
\inst{
$^{1}$Laboratoire des Solides Irradi\'{e}s, CNRS UMR 7642 \& 
CEA/DSM/DRECAM, \\
Ecole Polytechnique, 91128 Palaiseau, France \\
$^{2}$Department of Materials Science, Japan Atomic Energy Research 
Institute, Tokai-mura, Ibaraki 319-1195, Japan  \\
$^{3}$Kamerlingh Onnes Laboratorium, Leiden University, P.O. Box 9506, 
2300 RA Leiden, the Netherlands \\
$^{4}$Institut de Physique Nucl\'{e}aire, CNRS-IN2P3, 91406 Orsay, France}

\recdate{ }    

\abst{
    Using 30 MeV C$_{60}$ fullerene irradiation, we have produced 
    latent tracks of diameter 20 nm and length 200 nm,
    near the surface of single crystalline Bi$_{2}$Sr$_{2}$CaCu$_{2}$O$_{8+\delta}$. 
    A preliminary transmission electron microscopy study shows evidence for a very 
    high density of deposited energy, and the ejection of material from the track core in 
    very thin specimens. The latent tracks reveal themselves to be exceptionally strong pinning 
    centers for vortices in the superconducting mixed state. Both the critical current density 
    and magnetic irreversibility line are significantly enhanced. The irradiated crystals present 
    salient features of the $(B,T)$ phase diagram of vortex matter {\em both} of pristine crystals, 
    such as the first order vortex phase transition, {\em and} the exponential 
    Bose-glass line characteristic of heavy ion-irradiated crystals. We show that the latter is 
    manifestly independent of the pinning potential. 
    }

    \kword{ Vortex, flux pinning, layered superconductor, latent 
    tracks, columnar defects, phase diagram}

\begin{document}
\maketitle

\section{Introduction}

The recent discovery that a moderate amount of columnar strong pinning 
centers, induced by heavy ion irradiation,\cite{Konczykowski91III,Konczykowski92} does 
not qualitatively affect the $(B,T)$ phase diagram of the layered 
superconductor Bi$_{2}$Sr$_{2}$CaCu$_{2}$O$_{8+\delta}$ \cite{Banerjee03} has revived 
interest in disorder effects on the thermodynamics of superconductors 
in a magnetic field.\cite{DasGupta02,DasGupta03} In the absence of disorder, 
the phase diagram is characterized by a single First Order Transition 
(FOT)\cite{Zeldov95II} from an ordered 
vortex Bragg glass,\cite{Giamarchi96} with both topological long range order, 
and long range order of the superconducting phase, to a vortex liquid that 
lacks either.\cite{Fuchs97II} This low field ``vortex solid'' has the dynamic properties 
of a true superconductor, with diverging flux creep activation 
barriers \cite{Fuchs98II,Yacine04} and vanishing linear resitivity, 
while in the vortex liquid the superconductor obeys Ohm's law. 
Banerjee et al. showed that Bi$_{2}$Sr$_{2}$CaCu$_{2}$O$_{8+\delta}$ crystals
with a sufficiently small density of amorphous columnar defects still 
show a FOT to the vortex liquid.\cite{Banerjee03}
A companion study by Menghini {\em et al.}  demonstrated that in the same crystals, the vortex
solid is highly disordered.\cite{Menghini03} For fields below
the ``matching field'' $B_{\phi} = n_{d}\Phi_{0}$, where $n_{d}$ is the areal 
density of defects and $\Phi_{0} = h/2e$ is the flux quantum, vortices adapt 
themselves optimally to the defect positions and Bragg glass order is completely 
destroyed. Thus, translational symmetry of the low-field vortex 
lattice does not seem to be a prerequisite for the FOT. 

The irrelevance of Bragg glass order can be understood by the fact that vortex 
tilt is what limits thermal vortex excursions near the transition.\cite{Colson03}
The difference in free energy of the Bragg glass and the vortex liquid is, for 
sufficiently low fields, essentially determined  by the difference in vortex tilt
modes between the low field phase, with well-defined (disentangled) vortex lines, 
and the vortex liquid in which vortices are entangled. The inclusion of correlated
disorder unambiguously increases pancake vortex correlations along the field direction in
the vortex liquid.\cite{Beek95,Doyle96,Lopez96I,Kosugi97,Sato97,Colson04,Banerjee04}
Apparently, vortex fluctuations are also affected, as evidenced by the 
increase of the  FOT field\cite{Banerjee03} and its eventual transformation into a 
second order Bose-glass transition.\cite{Nelson93,Larkin95,vdBeek2001} 
Nevertheless, in the portion of the $(B,T)$ phase diagram that is 
transformed from vortex liquid to Bose glass by correlated disorder, 
the vortices show dynamical properties reminiscent of a two-dimensional (2D) 
system; also, the position of the Bose-glass to liquid transition in layered 
superconductors is well described if one considers the system to be 
essentially 2D with weak coupling between layers.\cite{vdBeek2001} The 
properties of the columnar defects, such as their size or density, were predicted 
to little affect the phase diagram of the irradiated superconductor, 
or the dynamical properties of the vortices.\cite{vdBeek2001}

Here, we report on the creation of large diameter ( 20 nm ) columnar 
defects in  optimally doped single crystalline Bi$_{2}$Sr$_{2}$CaCu$_{2}$O$_{8+\delta}$
by 30 MeV C$_{60}$ irradiation. A preliminary Transmission Electron 
Microscopy study confirms the presence of latent tracks, as well as 
their size distribution. The large track diameter is due to the very high energy density 
deposited by the fullerene fragments in the Bi$_{2}$Sr$_{2}$CaCu$_{2}$O$_{8+\delta}$ 
matrix, $S_{e} = 70$ keVnm$^{-1}$. The relatively small velocity of 
the C$_{60}$ fragments imply the tracks are very short (an estimated 200 nm) in 
comparison to the thickness of usual  Bi$_{2}$Sr$_{2}$CaCu$_{2}$O$_{8+\delta}$ 
single crystals. The defect potential experienced by vortex 
lines in the mixed state is thus intermediate between that of surface 
damage and columnar tracks. Magneto-optical observations of the flux 
density on both sides of a 40 $\mu$m thick single crystal are used to 
measure the critical current density and the current-voltage 
characteristics resulting from vortex pinning by the tracks. The creation
of a random distribution of very strong pinning centers at one crystal surface 
only is reminiscent of Ref.~\cite{Fasano03}. Those authors deposited an ordered 
array of moderately strong pinning Fe particles on one surface of clean 
Bi$_{2}$Sr$_{2}$CaCu$_{2}$O$_{8+\delta}$ single crystals of different thicknesses 
by means of Bitter decoration of the vortex Bragg glass. From their results, we expect 
vortex Bragg glass (translational) order to be destroyed only at the irradiated surface. 
Nevertheless, Differential Magneto-Optical (DMO) Imaging (see below) shows that the first 
order vortex lattice transition is preserved. Upon increasing temperature or field, the vortex 
solid transists, at high temprature, to a vortex liquid, and at lower temperature, to a pinned, 
presumably glassy phase. The temperature at which this glass transists 
to the vortex liquid  is the same as the Bose-glass temperature 
in heavy-ion irradiated Bi$_{2}$Sr$_{2}$CaCu$_{2}$O$_{8+\delta}$ 
crystals,\cite{vdBeek2001} demonstrating that the larger pinning energy expected for the 
C$_{60}$--tracks does not influence its position.

\section{Crystal growth and irradiation procedure}

Bi$_{2}$Sr$_{2}$CaCu$_{2}$O$_{8+\delta}$ single crystals were grown 
using the travelling-solvent floating zone method at 
the FOM- ALMOS center (University of Amsterdam, the Netherlands), in
200 mbar oxygen partial pressure.\cite{MingLi2002I} The crystals were extracted 
from the boule using a razor blade, and annealed in air for two weeks at 
800 $^{\circ}$C. A suitable specimen, of dimensions $980 \times 800 \times 20$ 
$\mu$m$^{2}$, without macroscopic defects such a grain 
boundaries or second-phase intergrowths, was selected for further 
measurements using magneto-optical imaging. Some of the crystal's 
physical properties, such as $T_{c} = 89$ K, characteristic for optimally doped 
Bi$_{2}$Sr$_{2}$CaCu$_{2}$O$_{8+\delta}$, the vortex lattice FOT 
field at low temperature (``second peak field''),\cite{Khaykovich96} 
$B_{FOT} = 300 \pm 10$ G, and the low temperature $I(V)$ characteristic, were 
also determined before the irradiation experiment. 

A number of other crystals from the same batch were finely ground to get 
thin fragments, and were placed on a 3 mm diameter copper grid covered with a 
very thin amorphous carbon film for Transmission Electron 
Microscopy (TEM) studies. All samples were subsequently irradiated 
with 30 MeV C$_{60}$ fullerene ions, at normal incidence and at room 
temperature, in the tandem accelerator of the  ``Institut de Physique 
Nucl\'{e}aire'' at Orsay University (Orsay, France). The samples were 
irradiated up to a fluence of  $1\times 10^{10}$ molecules ${\cdot \rm cm^{-2}}$. 
For the crystal used in the magneto-optical measurements, the track 
density, obtained from TEM corresponds to the C$_{60}$--fluence and 
to the matching field $B_{\phi} = 0.2$ T. After the irradiation, the crystal 
retained its initial critical temperature, $T_{c} = 89.0$ K.

\begin{figure}[t]
    \vspace{1mm}
    \centerline{ \epsfxsize 8cm \epsfbox{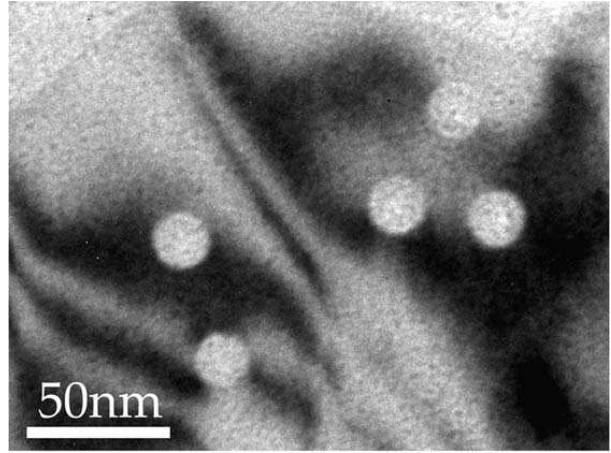} }
    \caption{Bright field image of tracks induced in 
    Bi$_{2}$Sr$_{2}$CaCu$_{2}$O$_{8+\delta}$ irradiated at normal incidence and at 300 K with 
    30 MeV C$_{60}$ fullerenes up to a fluence of $10^{10}$ cm$^{-2}$.}
    \label{Fig:tracks1}
    \end{figure}
   
\section{Transmission electron microscopy study}
\subsection{Experimental observations}

The samples destined for TEM studies were examined with a Philips CM 30 
transmission electron microscope (TEM) operating at 300 kV. Since tracks created 
in Bi$_{2}$Sr$_{2}$CaCu$_{2}$O$_{8+\delta}$ are rather unstable under electron
irradiation, the beam current in the TEM was maintained at a low level in order 
to minimize the change of track structure during observation. The energy losses 
and the range of  C$_{60}$ clusters were calculated using the TRIM code 
\cite{Biersack80}, supposing that  the energy loss of the cluster is the sum of 
the contributions of each constituent, {\em i.e.} here as sixty times the energy 
loss of individual carbon atoms of the same velocity. This additive rule is 
verified by a

\begin{figure}[b]
    \centerline{ \epsfxsize 8cm \epsfbox{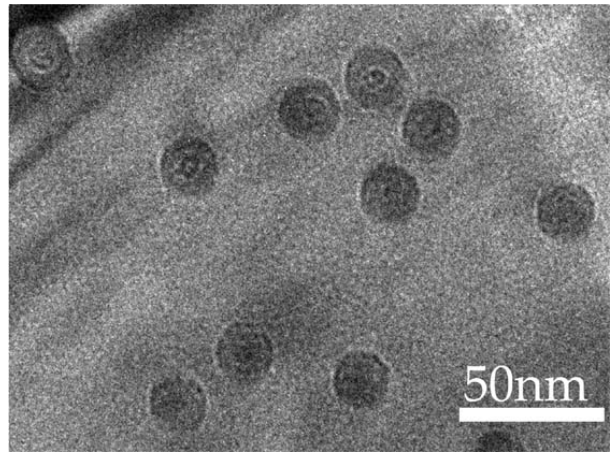} }
    \caption{Bright field image of tracks induced in Bi$_{2}$Sr$_{2}$CaCu$_{2}$O$_{8+\delta}$ 
	    irradiated at normal incidence and at 300 K with 30 MeV C$_{60}$ fullerenes up to a fluence 
	    of $10^{10}$ cm$^{-2}$ (no reflection is excited, slight defocus of the objective lens).}
	    \label{Fig:tracks2}
\end{figure}

\begin{figure}[t]
    \centerline{ \epsfxsize 8cm \epsfbox{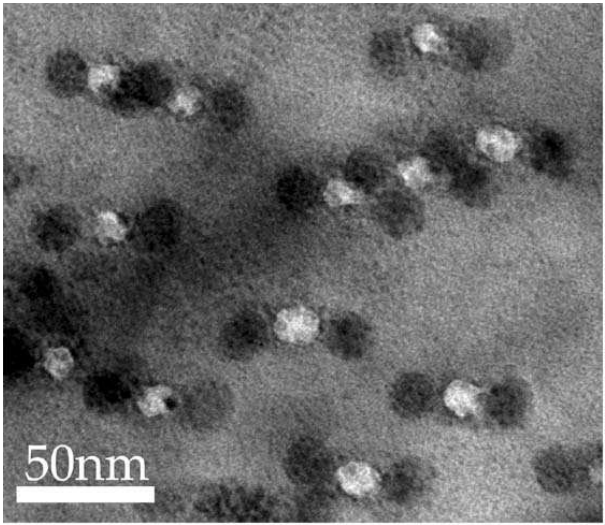} }
    \caption{Bright field image of tracks induced in very thin ($\leq 
    50$ nm) regions of Bi$_{2}$Sr$_{2}$CaCu$_{2}$O$_{8+\delta}$ irradiated at normal incidence 
    and at 300 K with 30 MeV C$_{60}$ fullerenes up to a fluence of 
    $10^{10}$ cm$^{-2}$. The sample was tilted in the microscope by 30$^{\circ}$.}
    \label{Fig:tracks3}
\end{figure}

\noindent number of experimental measurements of energy losses of various cluster
ions \cite{Baudin94,Tomaschko95,Benhamu97}. For 30 MeV fullerene ions 
slowing down in 
Bi$_{2}$Sr$_{2}$CaCu$_{2}$O$_{8+\delta}$ we find a  linear rate 
of energy  deposition  in electronic processes 
$S_{e} = (dE/dx)_{e} = 69.7$ keVnm$^{-1}$. The low incident
velocity of the cluster and the lateral straggling of the projectile 
constituents \cite{Dunlop97} lead to a rapid decrease of the 
linear  rate of energy 
deposition in electronic processes by spatially correlated fragments. 
A comparison with previous results in other materials\cite{Jensen98} 
suggests that the energy deposition could fall below the threshold for 
the creation of latent tracks at a depth of approximately  200 nm.

\begin{figure}[t]
    \centerline{ \epsfxsize 8cm \epsfbox{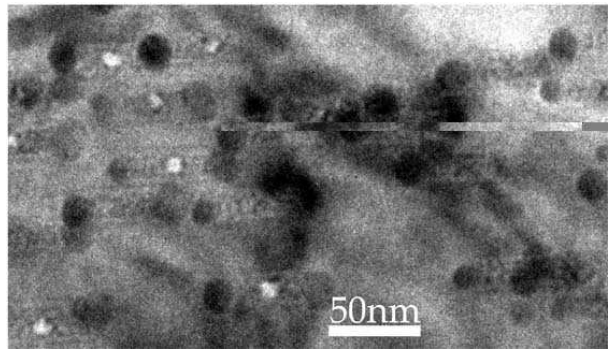} }
    \caption{Bright field image of tracks induced in Bi$_{2}$Sr$_{2}$CaCu$_{2}$O$_{8+\delta}$ 
    irradiated at normal incidence and at 300 K with 30 MeV C$_{60}$ fullerenes up to a fluence of
    $10^{10}$ cm$^{-2}$. The sample was tilted in the microscope by 
    30$^{\circ}$. The sample thickness increases from left to right. 
    No white contrast in the center of the tracks is observed in regions 
    thicker than $\approx 130$ nm.}
    \label{Fig:tracks4}
\end{figure}

The main result of this preliminary electron microscopy work is that large diameter 
tracks are observed. Figures \ref{Fig:tracks1} and \ref{Fig:tracks2}  show bright
field images of tracks in \linebreak Bi$_{2}$Sr$_{2}$CaCu$_{2}$O$_{8+\delta}$ irradiated 
with 30 MeV C$_{60}$ at normal incidence. In Fig.~\ref{Fig:tracks1} the cross-section of the 
tracks gives rise to a strong contrast. Namely, the regions located in the vicinity of the projectile
trajectories are highly damaged, so that they diffract very differently from the surrounding 
matrix and give uniform white contrasts. In Fig.~\ref{Fig:tracks2}, taken in different imaging 
conditions (no reflection is excited, slight defocus of the objective lens),  the contrast 
is not uniform: Fresnel fringes circle the regions corresponding to local variations of the
mass thickness of the target (local thickness multiplied by the density of the material 
\cite{Reimer97,Williams96}). We now find an outer contrast of the same diameter as that seen 
in Fig.~\ref{Fig:tracks1} and an inner contrast of a smaller diameter corresponding to the 
projection in the sample thickness of a different type of ``object'' located on the track axis.
Figures~\ref{Fig:tracks3} and \ref{Fig:tracks4} show bright field images of irradiated samples 
that have been tilted by 30$^{\circ}$ in the electron microscope relative to the incident ion 
direction. Figure~\ref{Fig:tracks3} was taken in a very thin sample region, close to the sample 
edge, and shows pairs of circular black contrasts located on both sample surfaces at the entrance 
and exit points of the C$_{60}$ projectiles. A circular white contrast is found between the areas 
of black contrast. Figure~\ref{Fig:tracks4} corresponds to a thicker region of the same sample 
(the sample thickness increases when going from the left to the right on Fig.~\ref{Fig:tracks4}), 
so that the projected distance between the two black contrasts is much larger than in 
Fig.~\ref{Fig:tracks3}. On the left part of Fig.~\ref{Fig:tracks4}, a circular white contrast 
is observed on the track axis approximately halfway between the projectile entrance and exit points. 
The diameter of the areas of white contrast is slightly smaller than the track diameter. On the right 
part of Fig.~\ref{Fig:tracks4}, corresponding to even thicker sample regions, the black contrasts at
both track ends are still visible, but the central white contrasts are no longer there. It is
difficult to define a precise thickness threshold for the creation of ÔÔwhiteÕÕ objects: the areas 
of white contrast are imaged in regions of thickness up to about 130 nm.
 
\subsection{Evolution of the track diameter}

\begin{figure}[t]
    \centerline{ \epsfxsize 8cm \epsfbox{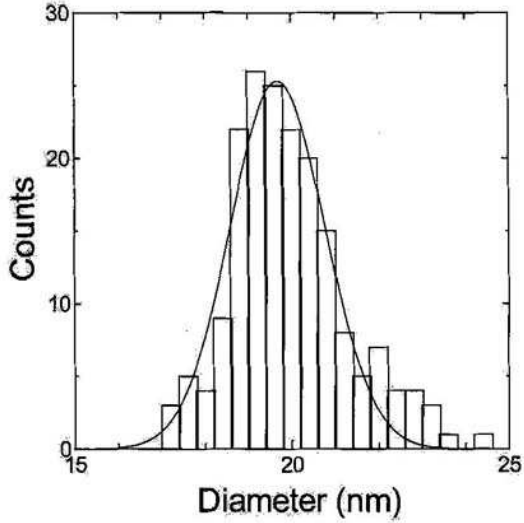} }
    \caption{Size distribution of tracks observed for Bi$_{2}$Sr$_{2}$CaCu$_{2}$O$_{8+\delta}$ 
    irradiated at 300 K with 30 MeV C$_{60}$ fullerenes.  
    The line in the figure is drawn following a Gaussian distribution.}
    \label{Fig:diameters}
\end{figure}

\begin{figure}[t]
    \centerline{ \epsfxsize 8cm \epsfbox{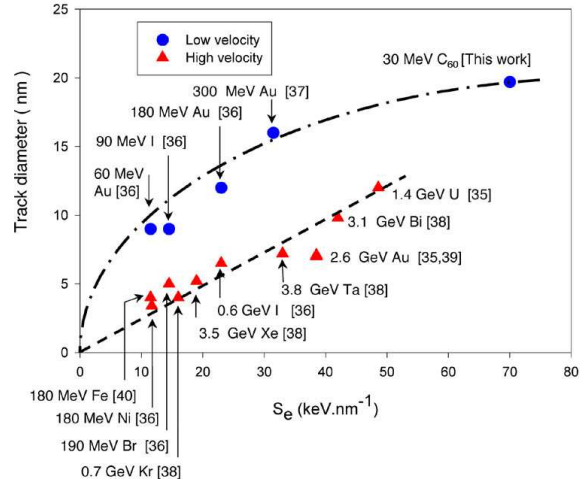} }
    \caption{Track diameters measured in Bi$_{2}$Sr$_{2}$CaCu$_{2}$O$_{8+\delta}$  irradiated with 
    monoatomic ions (from refs.~\protect\cite{Wiesner96,Sasase2001,Zhu93,Kuroda2001,Huang99,Huang2000})
    plotted as a function of the electronic stopping power $S_{E}$ for irradiations. The circles and 
    triangles respectively correspond to the low velocity and high velocity projectiles.  Our 
    relative to 30 MeV C$_{60}$ ions has been added (last point on the low velocity branch).}
    \label{Fig:tracksizes}
    \end{figure}

The measured diameters of the tracks generated in Bi$_{2}$Sr$_{2}$CaCu$_{2}$O$_{8+\delta}$ by 30 MeV 
C$_{60}$ fullerenes are plotted in Fig.~\ref{Fig:diameters}. The mean diameter is estimated to be 
$D=19.7$ nm supposing that the size distribution follows a Gaussian distribution.
Numerous observations by transmission electron microscopy of tracks created in 
Bi$_{2}$Sr$_{2}$CaCu$_{2}$O$_{8+\delta}$ by monoatomic energetic heavy ions (Cu to U, energies of 100 MeV
to 3 GeV) have been reported in the literature 
\cite{Wiesner96,Sasase2001,Zhu93,Kuroda2001,Huang99,Huang2000}. The tracks were found to be amorphous 
with diameters in the range $3-16$ nm. All these results have been plotted 
in Fig.~\ref{Fig:tracksizes}, which shows the evolution of the track diameter determined by TEM 
as a function of the linear rate of energy deposition in electronic 
processes $S_{e}$. The data points lie on two different branches relative to ``high velocity'' and 
``low velocity'' projectiles, referring to the fact that the corresponding ion velocity is associated
to a point respectively located above or below the maximum (Bragg peak) in the slowing-down curve. 
This fact that at the same $S_{e}$ the diameters for low-velocity mono-atomic ions are larger than 
those for high-velocity mono-atomic ions, called in the literature the ``velocity effect'', was already 
observed in various target types \cite{Meftah93,Jensen98II,Wang96}. These result confirm once more that 
not only $S_{e}$ but also the ion velocity must be taken into account to describe track formation. 
30 MeV fullerene ions deposit a high rate of energy in electronic processes during their slowing-down 
due to the addition of the effect of each atom in the cluster. However, as they are slow projectiles 
($v/c\approx 0.01$), the energy of the ejected $\delta$-electrons is very low, so that the energy 
is deposited in a very close vicinity of the projectile path, leading to very high energy densities. 
Thus very strong structural modifications are expected. In Fig.~\ref{Fig:tracksizes}, the diameter 
associated to cluster ions is higher than any measured using monoatomic ions and lies on the 
``low velocity'' branch as could be expected.

\subsection{Ejection of material from the track core}

The white contrast areas (Figs.~\ref{Fig:tracks3} and \ref{Fig:tracks4}) correspond to regions of
mass thickness lower than that of the surrounding matter, whereas the areas of black contrast most 
likely correspond to regions that are locally thicker. The formation of such features can be explained
as follows. Just after the passage of the projectile, the excited energy density around the projectile
path is very high.  The excess energy can relax either radially or along the track axis.  If the sample
is thin, {\em i.e.} if both surfaces are not too far from the center of the track, matter can be expelled
along the track axis and ejected.  This leads to the creation of regions of very low atomic density in
the center of the track, which correspond to the white contrast.  If the sample thickness is too
large, this process cannot occur.  Some of the expelled matter stays at both track ends and forms 
locally thicker regions, which are subsequently observed as black contrasts at the entrance and the exit
points of the projectile. As far as we know, similar areas of white contrast inside a damage track have 
never been reported for monoatomic ion irradiation. Note that a very similar 
result has previously been observed in insulating Y$_{3}$Fe$_{5}$O$_{12}$ garnet
\cite{Jensen98II,Jensen98}. Spherical white contrast areas were  observed in thin regions 
of the garnet sample after irradiation with cluster ions (10-40 MeV 
C$_{60}$), but could never be observed after monoatomic ion irradiation. Therefore, it is likely that 
the creation of the areas of white contrast, presumably corresponding to areas of lower density, 
inside damage tracks is a general phenomenon originating from the very high density of energy deposited 
in poorly conducting materials. The level of excitation is so high that the resulting pressure can indeed
eject material from the center of the track as long as the surfaces are not too far away.

\section{Magneto-optical imaging of the flux distribution in the superconducting mixed state}
\subsection{Experimental procedure}

The critical current density and magnetic relaxation experiments were carried out using  
Magneto-Optical Imaging\cite{Dorosinskii92} at temperatures between 
20 and 60 K, and inductions up to $H_{max} = 700$ G. In this technique, a 4 
$\mu$m--thick ferrimagnetic garnet indicator film with in-plane 
anisotropy, covered by an Al mirror, is placed on the Bi$_{2}$Sr$_{2}$CaCu$_{2}$O$_{8+\delta}$ 
crystal surface. Linearly polarized light is transmitted through 
the garnet, reflected on the Al mirror, and transmitted through the 
garnet a second time. The Faraday rotation of the light's 
polarization vector, which is proportional to the perpendicular 
component of the garnet magnetization, and to the local induction 
at the Bi$_{2}$Sr$_{2}$CaCu$_{2}$O$_{8+\delta}$ crystal surface, is 
observed using a nearly crossed analyzer. The resulting image, 
recorded using a polarized light microscope and a CCD camera, reveals the modulation 
of the perpendicular ($z$) component of the local induction at the sample surface 
as intensity differences.

\begin{figure}[t]
    \centerline{ \epsfxsize 8cm \epsfbox{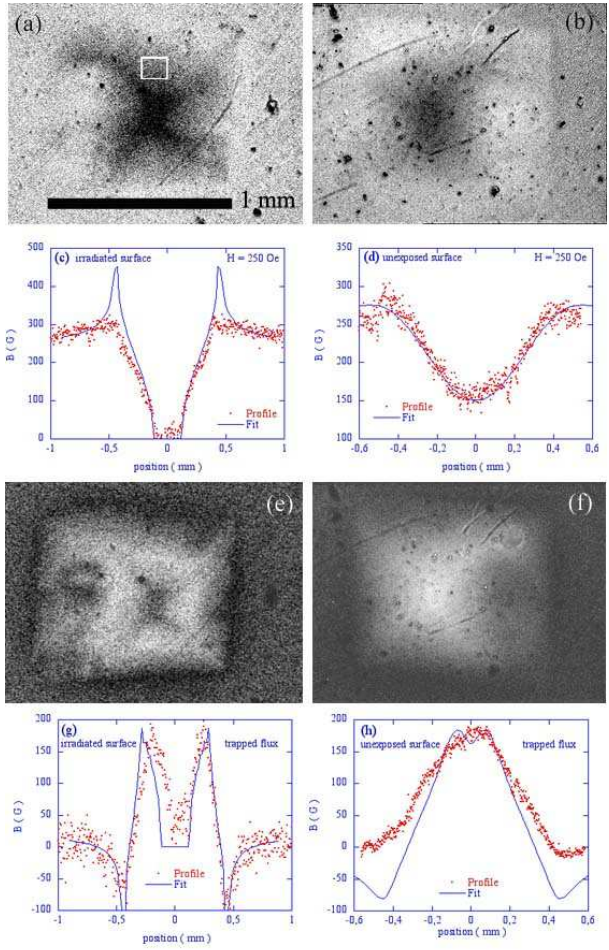} }
    \caption{ 
    Magneto-optical images of the magnetic induction on the 
    surface of the Bi$_{2}$Sr$_{2}$CaCu$_{2}$O$_{8+\delta}$ 
    crystal, after irradiation with 30 MeV C$_{60}$. The images show 
    the flux density on the crystal surface that was exposed to the 
    molecular beam (a,e), as well as on the unexposed surface (b,f).
    (a,b) Flux penetration at 45.6 K, after zero-field 
    cooling and the application of an external field, $H_{a} = 250$ 
    Oe. The small rectangle denotes the area in which the data of 
    fig.~\protect\ref{fig:hysteresis} were obtained. (c,d) Flux 
    profiles taken from the top to the bottom of images (a) and (b), 
    respectively, through the crystal center. Drawn lines show the 
    flux profile expected from the critical state model \protect\cite{Brandt93}, 
    supposing that the critical current of $1.6 \times 10^{11}$ 
    Am$^{-2}$ flows only in the 200 nm--thick top layer containing 
    the columnar tracks induced by the C$_{60}$--irradiation; (c)  at 
    1$\mu$m above the surface (d) at 60 $\mu$m from the irradiated surface. 
    (e,f) Trapped flux after removal of a maximum applied field of 320 
    Oe (e) and 580 Oe (f). (g) and (h) remanent flux profiles expected from 
    the critical state,  at 1$\mu$m above the surface (g) and at 60 $\mu$m from 
    the irradiated surface (h). 
    }
    \label{fig:MO} 
\end{figure}

Figure~\ref{fig:MO}(a) shows an image of the 
irradiated crystal after zero-field cooling to 45.6 K and the 
application of a 250 Oe external field. The indicator film was placed

\noindent directly on the irradiated surface. Bright areas reflect a high 
local induction, while dark areas have small local induction.  All images 
were calibrated using the light intensity measured far from the 
sample during a field sweep. Figure~\ref{fig:MO}(a) clearly shows a flux density distribution 
that is characteristic of the Bean model and strong flux pinning,
\cite{Bean62,Brandt93,Zeldov94,Brandt96II} at a temperature and field at which 
unirradiated crystals show no bulk flux pinning.\cite{Indenbom94IV,Zeldov94II}  
Figure~\ref{fig:MO}(c) shows a flux profile, taken from the top to the 
bottom of fig.~\ref{fig:MO}(a), through the crystal center. The flux 
profile is compared to that expected over a thin strip sample of 
width $w$ and thickness $d$ in the critical state. This is obtained by appling the Biot-Savart 
law to the current density distribution derived in ref.~\cite{Brandt93},
\begin{eqnarray}
    j &= & j_{c}   \hspace{3.6cm} ( x_{f} < |x| < w ) \nonumber  \\
    j &= &  \frac{2}{\pi} j_{c} \arctan Q  \hspace{2cm} ( 0 < |x| < x_{f} )  \nonumber
    \end{eqnarray}
    
\noindent where $Q = x [1-(x_{f}/d)^{2}]^{1/2}(x_{f}^{2}-x^{2})^{-1/2}$, $H_{a}$ 
is the applied field, and 
\begin{equation}
  x_{f} = \frac{w}{2} \frac{1}{\cosh\left(\pi H_{a} / jd \right)}
    \label{eq:fluxfront}
\end{equation}   

\noindent is the position of the flux front. Good agreement is obtained if we 
assume that the indicator is separated from the sample by a distance 
of 1 $\mu$m, and that a (field-independent) critical screening current of 
$1.6 \times 10^{11}$ 
Am$^{-2}$ flows only in the top 200 nm of the crystal. This layer 
corresponds to that which contains the tracks created by the 
C$_{60}$--irradiation. Figure~\ref{fig:MO}(e) shows 
the trapped flux on the irradiated crystal surface after removal of the applied 
field. In this case, good agreement is again obtained by assuming that 
the critical current flows only in the layer containing the tracks, 
and applying the rule $B(x,H_{a}) = B(x,H_{max}) - 2 
B(x,\frac{1}{2}H_{max}-\frac{1}{2}H_{a})$ applicable following a 
reversal of the field sweep direction. In contrast, images collected with the 
magneto-optical indicator film placed on the opposite surface, that 
remained unexposed to the C$_{60}$ beam, show much smoother, shallower 
flux profiles. These show that a large distance, exceeding the sample thickness, 
separates the current-carrying layer containing the tracks from the indicator surface.
Thus, we surmise that the critical current flows mainly in the 200 nm 
thick layer containing the tracks,

\begin{figure}[h]
    \centerline{\epsfxsize 6.5cm \epsfbox{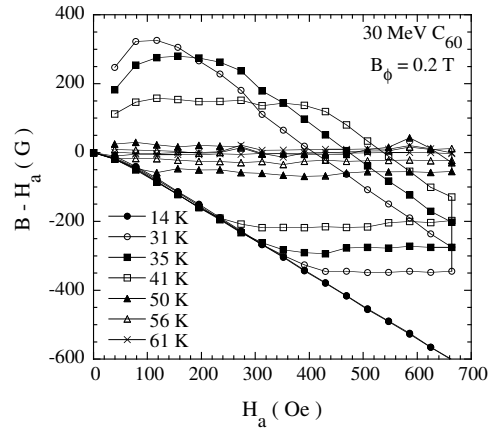}}
    \caption{Hysteresis of the ``self-field'', defined as the 
    difference between the local induction $B$ and the applied field 
    $H_{a}$, averaged on the square region of the 
    C$_{60}$--irradiated Bi$_{2}$Sr$_{2}$CaCu$_{2}$O$_{8+\delta}$ crystal 
    outlined in fig.~\protect\ref{fig:MO}. Data are for images taken 
    on the irradiated surface.}
    \label{fig:hysteresis}
\end{figure}

\noindent and that all screening in 
the rest of the crystal is produced by this current only. 

\subsection{Determination of the critical current density}%

\begin{figure}[t]
    \vspace{2mm}
   \centerline{\epsfxsize 8cm \epsfbox{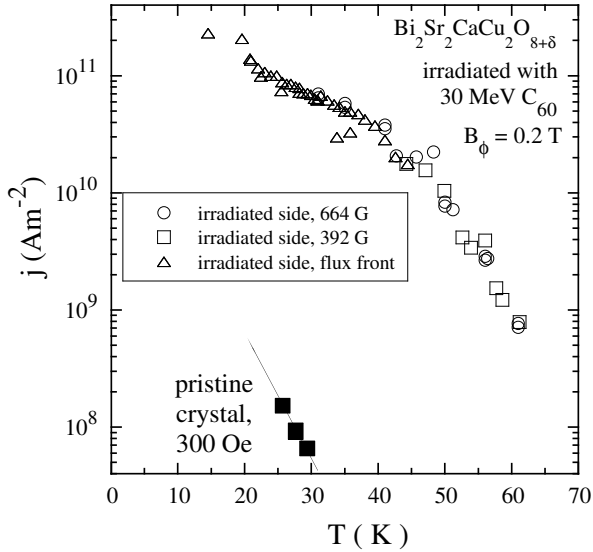}}
    \caption{ Sustainable current density, at $E = 6$ $\mu$Vm$^{-1}$, 
    of the C$_{60}$-irradiated Bi$_{2}$Sr$_{2}$CaCu$_{2}$O$_{8+\delta}$ 
    single crystal. The (field-independent) current was measured from 
    magneto-optical images of the flux distribution on 
    the irradiated surface. Induction values are 664 G ($\circ$), 
    392 G (\protect\raisebox{1mm}{\protect\framebox{\hspace{1pt}}}), and 0 
    (position of the flux-front) ($\triangle$). Filled squares joined by a 
    drawn line represent data on the crystal prior to irradiation.}
    \label{fig:j-vs-T}
\end{figure}

Magnetic hysteresis loops of the local induction were 
recorded by sweeping the external field at a rate of  120 Oe/s, corresponding to an 
induced electric field at the crystal edge,  $E = 6$ $\mu$Vm$^{-1}$. This was 
achieved by controlling the bipolar power supply output with a 0.1 Hz triangular 
wave. A synchronized TTL signal triggered image acquisition at the 
rate of 4 images / s. For these experiments, the polarizer/ analyzer 
pair of the microscope was uncrossed by 10$^{\circ}$. After conversion of 
the image intensity to flux density, hysteresis loops of the local 
induction $B(H_{a})$ were determined in several points on the crystal 
surface. Subtraction of the applied field yielded hysteresis loops of 
the crystal ``self--field'' $B-H_{a}$, such as depicted in 
fig.~\ref{fig:hysteresis}. The sustainable current density was 
obtained from the critical state model.\cite{Brandt93,Zeldov94,Brandt96II} At fields 
above the full penetration field $H_{p}$ (at which the flux front has penetrated
to the center of the crystal) and less than $H_{max}-H_{p}$ (at which 
the remagnetization front has reached the crystal center),
the width $\Delta (B-H_{a})$ of the hysteresis loops, which is proportional 
to the sustainable current density $j$, turns out to be field--independent. 
The absolute magnitude of the current density was verified using the 
relation (\ref{eq:fluxfront}) between applied field and the position of 
the flux front,\cite{Brandt93} taking the crystal width $w = 800$ 
$\mu$m and $d = 200$ nm, the thickness of the surface layer containing continuous 
 latent tracks. In this manner, we obtained the temperature dependence of $j$ on  
the irradiated surface, between 20 and 60 K, for 
inductions up to 664 G (fig.~\ref{fig:j-vs-T}). 

\begin{figure}[t]
       \centerline{\epsfxsize 8cm \epsfbox{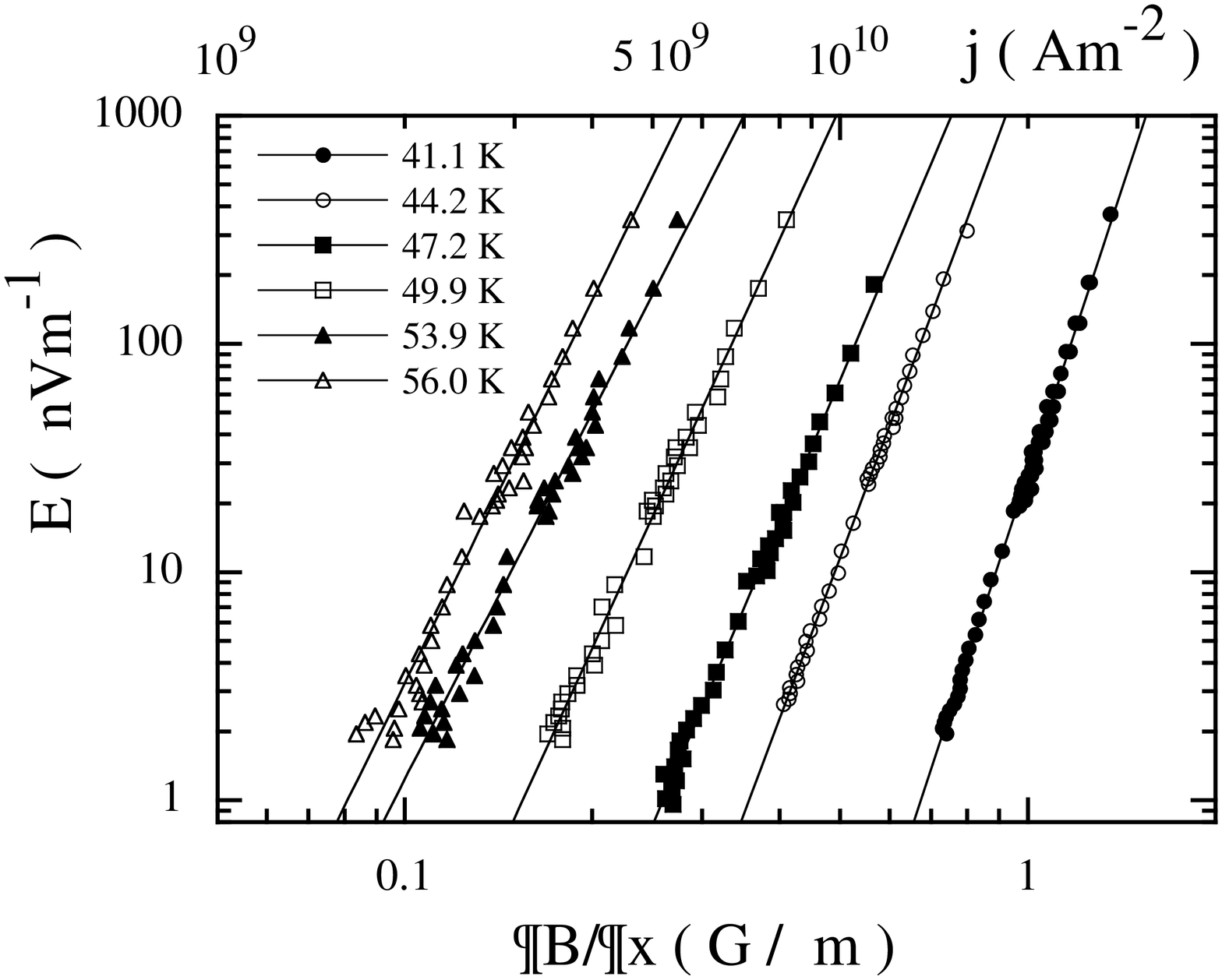}}
    \caption{Power-law $I(V)$--characteristics, measured by 
    magneto-optical imaging of the relaxation of the non-equilibrium 
    magnetic induction distribution on the  C$_{60}$-irradiated surface of the 
    Bi$_{2}$Sr$_{2}$CaCu$_{2}$O$_{8+\delta}$ single crystal.} 
    \label{fig:IV}
\end{figure}

\subsection{Magnetic relaxation and $I(V)$--curves}

Experiments on the temporal relaxation of the distribution of trapped flux were 
prepared by zero--field cooling to the working temperature and applying a 
field of 650 Oe. A two-channel synthesizer was used to simultaneously 
trigger the decrease of the applied field to 50 Oe, and the acquisition 
of magneto-optical images of the trapped flux. Images were acquired 
between 1 s and 200 s after reaching the 50 Oe target field. The 
flux distributions always corresponded to the Bean critical state 
profile, in accordance with a strongly nonlinear $I(V)$--characteristic. 
From the relaxing profiles, we obtained the current density as the 
induction gradient, $j = - \partial B_{z} / \partial x$, and the electric 
field as $E = \int_{0}^{w/2} (\partial B /\partial t) 
dx$.\cite{Abulafia95} The electric field was verified to decay as $E \propto 
1/t$.\cite{Gurevich94} The process was repeated for a number of temperatures 
at which full flux penetration could be achieved, with 
images taken both on the irradiated and on the unirradiated side of the 
crystal. Resulting $I(V)$-curves are displayed in fig.~\ref{fig:IV}. 
The $I(V)$--curves follow a power-law at all investigated 
temperatures.  The long-time 
behavior of the electric field at the sample edge, which for the measured power-law 
$I(V)$--relation follows $E(w/2) = \frac{1}{4}\mu_{0}j_{c}w^{2} 
(\tau_{0} + t)^{-1}$, with $\tau_{0} \approx 10$ ms a transient time, allows an 
estimate of the ``true'' critical current density 
$j_{c} = t \times [4 E(w/2, t \gg \tau_{0})/\mu_{0}w^{2}]$. This was 
found to coincide, within experimental error, with the data points of 
fig.~\ref{fig:j-vs-T}.

\subsection{Determination of the Phase Diagram}

\begin{figure}
    \centerline{\epsfxsize 7cm \epsfbox{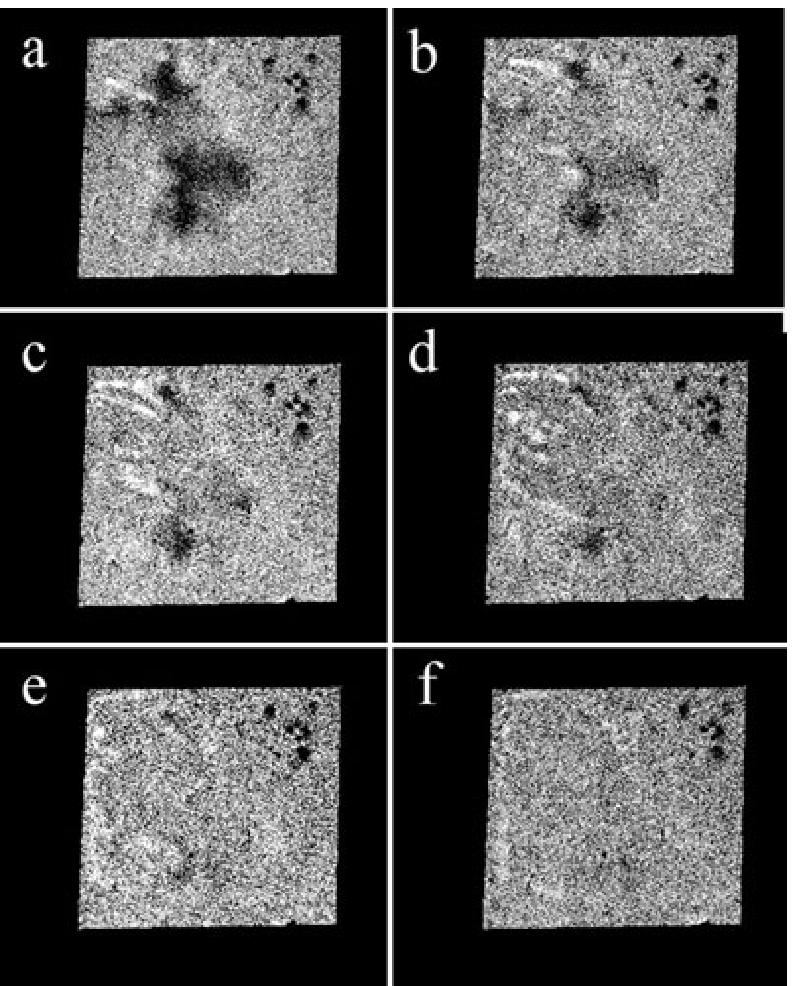} }
    \centerline{\epsfxsize 7cm \epsfbox{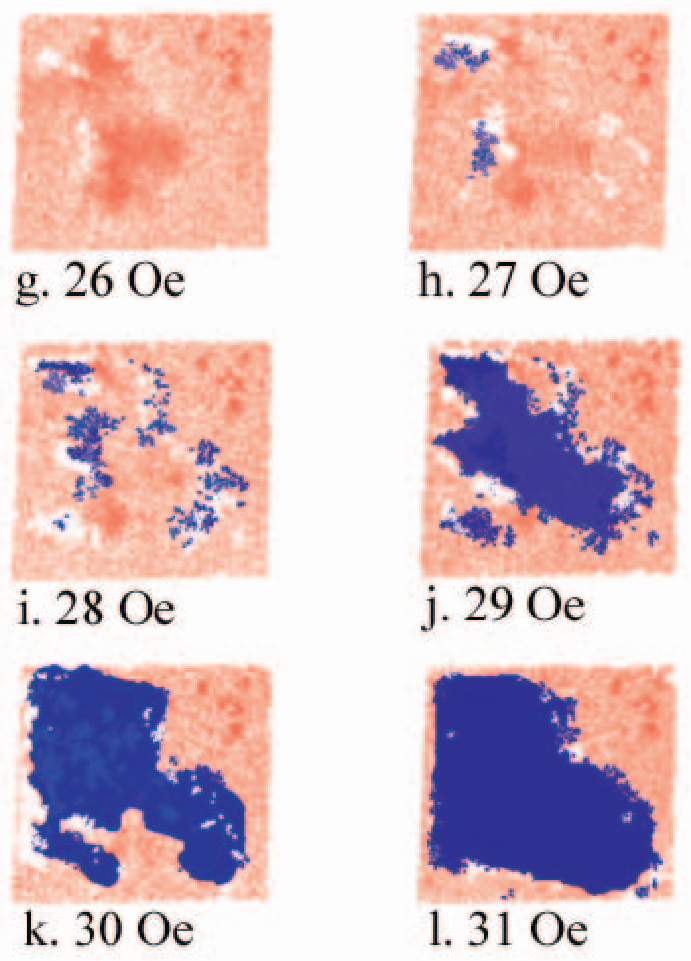} }
    \caption{Differential magneto-optical (DMO) images of the first order 
    vortex phase transition in an optimally doped
    Bi$_{2}$Sr$_{2}$CaCu$_{2}$O$_{8+\delta}$ single crystal irradiated 
    with 30 MeV C$_{60}$ fullerenes. The frames follow the crystal 
    outline. The images were taken at $T = 
    82.0$ K, in applied magnetic fields of 26 (a) to 31 Oe (f), each 
    image corresponding to a 1 Oe step. The field modulation $\Delta H_{a}$ 
    was 0.5 Oe. Dark areas show diamagnetic screening, while white 
    areas show the paramagnetic response due to the progression of 
    the phase trasnformation front. (g)-(l) Interpretation of images 
    (a)-(f) in terms of the growth of the vortex liquid phase (blue) 
    at the expense of the vortex solid (red).}
    \label{fig:DMO-images}
\end{figure}

\begin{figure}[t]
    \centerline{\epsfxsize 8cm \epsfbox{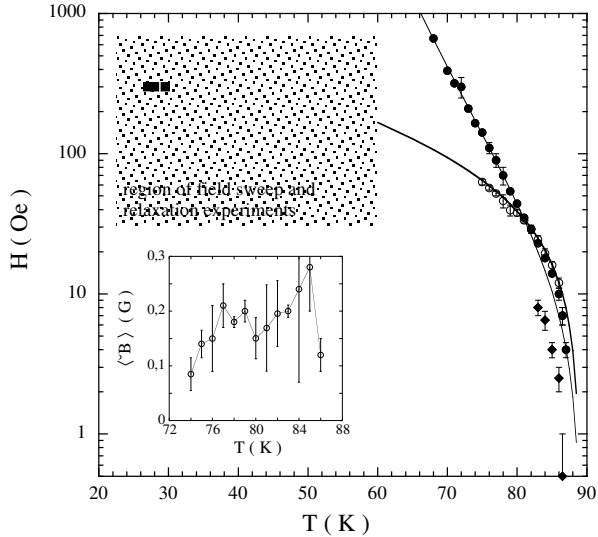} }
    \caption{Phase diagram of single crystal 
    Bi$_{2}$Sr$_{2}$CaCu$_{2}$O$_{8+\delta}$ irradiated 
    with 30 MeV C$_{60}$ fullerenes. Symbols denote the field of first 
    flux penetration ($\diamond$), the First Order 
    Transition field $H_{FOT}$ ($\circ$), and the irreversibility field $H_{irr}$ 
    ($\bullet$) after C$_{60}$ irradiation. The filled squares denote 
    the field of the ``second-peak'' transition, or low temperature part of the 
    FOT, in the pristine crystal.\cite{Yacine04,Avraham2001}
    Error bars denote the spread of FOT fields within the crystal.
    The drawn line through the FOT data is a guide to the eye, the 
    drawn line through the $H_{irr}$ data is a fit to 
    eq.~(\protect\ref{eq:Feigelman}) with parameters $B_{cr} = 
    70$ G and $\varepsilon_{0}(0)s/k_{B} = 770$ K. Inset: magnitude of 
    the equilibrium change in flux density at the FOT,  $\Delta B_{FOT}$. The error bars 
    show the spread of $\Delta B_{FOT}$ within the sample.}
    \label{fig:phase-diagram}
\end{figure}

Differential Magneto-Optical Imaging  \cite{Soibel2000,Soibel01} was 
used to determine the $(B,T)$ phase diagram. After stabilization of 
the temperature, ten  magneto-optical images are acquired at the target 
applied field $H_{a}$, and averaged. The magnetic field is then 
increased by $\Delta H_{a} = 0.5$ Oe, whence ten new images are acquired 
and consecutively subtracted from the initial average, to yield a 
differential image. The process is repeated twenty times to increase 
the signal-to-noise ratio. The result is an image of the ``local 
permeability'', $\Delta B / \Delta H_{a}$ (see fig.~\ref{fig:DMO-images}). 
Areas of $\Delta B / \Delta H_{a} = 1$, such as the space surrounding the 
crystal, would show up as grey (not shown in 
fig.~\ref{fig:DMO-images} that displays only the surface within the crystal 
perimeter). Regions that show diamagnetic screening, such 
as the crystal center in fig.~\ref{fig:DMO-images}(a), are revealed 
as black. Areas that, in the field interval between $H_{a}$ and 
$H_{a} + \Delta H_{a}$, undergo the phase change from vortex solid to 
vortex liquid, experience an increase $\Delta B_{FOT}$ in the equilibrium flux (vortex) 
density and thus have positive 
permeability.\cite{Soibel2000,Morozov96II} Therefore, the clear white 
areas show the progression of the phase transformation front in the 
field interval ($H_{a}$, $H_{a} + \Delta H_{a}$). The magnitude of 
the equilibrium flux density change $\Delta B_{FOT}$ is plotted versus 
temperature in the Inset of fig.~\ref{fig:phase-diagram}. 
Figure~\ref{fig:DMO-images} shows a series of images at 82.0 K, as the 
applied field is increased from 26 Oe to 31 Oe in 1 Oe steps. One 
sees that the vortex liquid nucleates in a number of spots in the 
upper left hand corner, rapidly expands, and finally occupies the 
whole crystal, except for a narrow ring around the crystal outer edge.

From the DMO images, we determine the phase diagram of the 
C$_{60}$--irradiated  Bi$_{2}$Sr$_{2}$CaCu$_{2}$O$_{8+\delta}$ single 
crystal. The field of first flux penetration corresponds to the applied 
field at which a non-zero flux can be observed at any spot in the crystal.
The FOT field is determined, for a given 
spot on the crystal surface, as the applied field at which the 
modulation $\Delta H_{a}$ produces a paramagnetic signal (white in 
fig.~\ref{fig:DMO-images}) at that location.  The local irreversibility 
line (IRL), or $H_{irr}(T)$ line at which the critical current density vanishes, is 
determined as the applied field at which the image intensity in a 
given spot coincides with the intensity caused by the full 0.5 Oe 
field modulation. Both the FOT line and the IRL depend on position, 
because of field inhomogeneity caused by both sample geometry 
and composition. The spread of $H_{FOT}$ and $H_{irr}$ is 
denoted by the error bars in fig.~\ref{fig:phase-diagram}. It is 
remarkable that, for high temperatures, the vortex solid undergoes the 
transition to the vortex liquid in the peripheral crystal areas of high local 
induction [white stripes near the crystal boundary in 
fig.~\ref{fig:DMO-images} (a,b)], while in the vortex solid near the crystal center vortices 
are still pinned and the field modulation is completely screened 
[black spots near the crystal boundary in fig.~\ref{fig:DMO-images} (a,b)]. Similar behavior was recently observed in inhomogeneous 
underdoped  Bi$_{2}$Sr$_{2}$CaCu$_{2}$O$_{8+\delta}$ 
crystals.\cite{vdBeek2003} Below 74 K, the equilibrium flux density change associated with the 
FOT becomes indistinguishable from the demagnetizing field due to 
strong screening currents in the sample center. Therefore, the FOT 
line is not plotted below this temperature.

\section{Discussion}

The $(B,T)$ phase diagram of the C$_{60}$ irradiated \linebreak
Bi$_{2}$Sr$_{2}$CaCu$_{2}$O$_{8+\delta}$ crystal surprisingly
features \em both \rm the FOT normally observed only in 
pristine\cite{Zeldov95II} or very lightly irradiated\cite{Banerjee03} single 
crystals, \em as well as \rm the exponential 
irreversibility line associated with the Bose-glass to vortex liquid 
transition in heavy-ion irradiated samples.\cite{vdBeek2001} Moreover, 
the two transitions occur at \em exactly the same locations \rm as in pristine 
and heavy-ion irradiated crystals, respectively.\cite{Zeldov95II,vdBeek2001} 
By virtue of their larger diameter $D = 20$ nm, one would expect the vortex
pinning energy per unit length of the latent tracks created by the C$_{60}$ 
fragments, $U_{p} \approx \varepsilon_{0}(D / 2 \xi)^{2}$ to be one order of
magnitude larger than that of tracks created by swift heavy ions, with $D = 7$ nm.
\cite{Drost99} Here 
$\varepsilon_{0}(T) = \Phi_{0}^{2}/4\pi\mu_{0}\lambda_{ab}^{2}(T)$ is the typical 
vortex energy, $\lambda_{ab}(T)$ is the in-plane penetration depth, and $\xi$ 
is the coherence length. We conclude that the irreversibility line 
in layered superconductors containing  columnar tracks is
not influenced by the pinning energy of the defect.

We now discuss the simultaneous observation of the FOT and the 
exponential glass-to-liquid irreversibility line. Plausibly, the latter 
line is determined only by the  outer layer containing the latent tracks,
while the FOT occurs only in the rest of the crystal. The large 
screening current in the irradiated layer of the crystal is 
sufficently strong to prevent flux penetration throughout the crystal, 
one can therefore observe its effects up to the irreversibility line 
on both crystal surfaces. As for the FOT, one might naively expect that 
this occurs only in the ordered Bragg glass, but not in the strongly pinned 
``top'' part of the vortex ensemble that occupies the latent tracks. Recent 
measurements by Banerjee {\em et al.} have shown that the FOT persists in 
crystals with a modest track density, $n_{d} < 5 \times 10^{8}$ cm$^{-2}$ or 
$B_{\phi} \leq 100$ G, even though the vortex solid in such crystals 
is completely disordered.\cite{Banerjee03,Menghini03} Here, we report 
the unprecedented observation of the 
FOT of the vortex solid in the presence of a track density $n_{d} = 1 
\times 10^{10}$ cm$^{-2}$, or $B_{\phi} = 0.2$ T. One possibility is 
that the vortex FOT is occuring only in the defect-free part of the 
crystal, and that the vortex density change at the FOT is propagated 
in the pinned upper portion of the vortex ensemble. It would imply 
that above 74 K, the lowest temperature at which we unambiguously 
observe the FOT, the sustainable current density  is smaller than the
$1.6 \times 10^{6}$ Am$^{-2}$ needed to screen $\Delta B_{FOT}$. An 
extrapolation of the high temperature part of the $j(T)$--curve of 
fig.~\ref{fig:j-vs-T} shows that, in spite of its very fast temporal 
relaxation, the screening current at 74 K is still more than an order 
or magnitude larger than this. Thus, in spite of the presence of very strong 
pinning centers, the vortex system still seems to undergo the FOT in the 
high temperature part of the phase diagram. 

\begin{figure}[t]
    \centerline{\epsfxsize 8cm \epsfbox{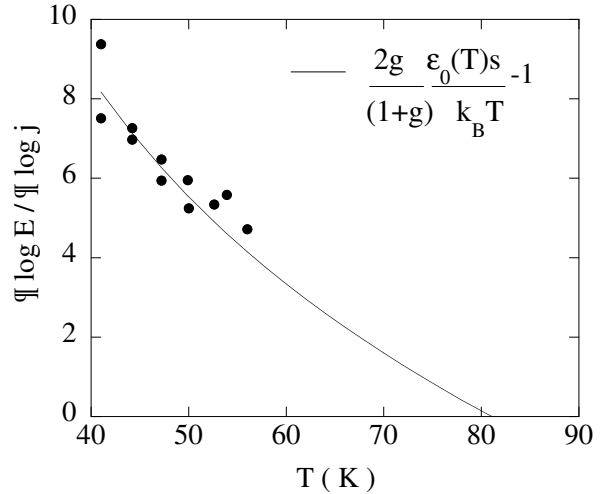}}
    \caption{Logarithmic slope $\partial \log E / \partial \log j$ of the current-voltage 
    characteristics of the  C$_{60}$--irradiated 
    Bi$_{2}$Sr$_{2}$CaCu$_{2}$O$_{8+\delta}$ single crystal. The drawn 
    line corresponds to the Coulomb gas\protect\cite{Minnhagen95} prediction $2 (g/1+g) 
    \varepsilon_{0}(T)s / k_{B}T - 1$ (with $\varepsilon_{0}(0)s / k_{B} = 
    770$ K, and $g = 0.4$).}
    \label{fig:power}
\end{figure}

We now turn to the nature of the high--field vortex state, at fields 
higher than $H_{FOT}$. Figure~\ref{fig:phase-diagram} shows that, at 
least in the layer containing  latent tracks, a 
very considerable portion of the vortex liquid (the region above the 
FOT line, without critical current before irradiation) 
is transformed to a pinned state of localized vortices, with a very 
high critical current density (fig.~\ref{fig:j-vs-T}).  
A framework that provides a conceptual language by which 
the behavior in the high field vortex state can be understood is the 
vacancy interstitial--unbinding transition of the 2D pancake vortex lattices 
contained within each superconducting layer. This transition is of 
Kosterlitz-Thouless type.\cite{Feigelman90II,Dodgson99} In 
a clean layered superconductor, the energy to create a 
vacancy-interstitial (dislocation quartet) in the pancake vortex lattice  
is non-zero only because of the non-zero interlayer coupling, which provides 
an effective ``substrate potential'' for the vortices. For weak (electromagnetic) coupling, 
it means the unbinding transition lies at a rather low temperature, $T \approx  
(g/1+g ) \varepsilon_{0}s/2 k_{B}$ $s = 1.5$ nm is the CuO$_{2}$ layer spacing.\cite{Dodgson99} 
The factor $g$ describes the relative weight of in-plane-- to interlayer pancake 
vortex interactions: if the first dominate, $g \ll 1$, if the 
interlayer coupling is important, $g > 1$.\cite{Dodgson99} 
In Ref.~\cite{vdBeek2001}, it was implicitly supposed that the introduction of columnar 
defects by heavy-ion irradiation effectively enhances $g$, by two 
mechanisms (i) the localization of vortices on the tracks prevents the
elastic screening of a vortex-lattice vacancy or interstitial (ii) the 
columns may artificially enhance the interlayer coupling by aligning the 
pancakes. The consequence is that the defect-pair energy is close to 
$\varepsilon_{0}s$, and that the irreversibility line, which is supposed 
to coincide with the vacancy--interstitial unbinding transition, does not 
depend on details of the pinning potential. 

Henceforth, the irreversibility line was estimated as \cite{vdBeek2001,Feigelman90II}

\begin{equation} 
    B_{irr} = B_{cr} \frac{\varepsilon_{0}(T)s}{k_{B}T}
        \exp \left[ \frac{\varepsilon_{0}(T)s}{k_{B}T}
    \right] , 
    \label{eq:Feigelman}
\end{equation}

with $B_{cr} \equiv \Phi_{0}/(\gamma s)^{2}$ the crossover 
field,\cite{Feigelman90II,Dodgson99} $\gamma s$ the 
Josephson length, and $\gamma$ the anisotropy parameter. Equation~(\ref{eq:Feigelman}) 
very well describes the irreversibility (Bose-glass) line in heavy-ion irradiated 
Bi$_{2}$Sr$_{2}$CaCu$_{2}$O$_{8+\delta}$ of different doping 
levels.\cite{vdBeek2001}. Figure~\ref{fig:phase-diagram} shows that such 
a comparison holds equally well for C$_{60}$--irradiated 
Bi$_{2}$Sr$_{2}$CaCu$_{2}$O$_{8+\delta}$. The fit to the 
irreversibility line yields $B_{cr} = 70$ G and 
$\varepsilon_{0}(0)s / k_{B} = 770$ K, consistent with $\lambda_{ab}(0) = 185$ nm
( {\em i.e.} a London penetration depth $\lambda_{L}(0) = 250$ nm) and $\gamma = 360$. 
These are the literature values for optimally doped 
Bi$_{2}$Sr$_{2}$CaCu$_{2}$O$_{8+\delta}$.\cite{Li97}
Thus, the transition from pinned vortex state to vortex liquid in C$_{60}$--irradiated 
Bi$_{2}$Sr$_{2}$CaCu$_{2}$O$_{8+\delta}$ bears all the hallmarks of the proposed 
vortex-unbinding transition: it obeys eq.~(\ref{eq:Feigelman}) and is independent of 
the pinning potential.

In case the high--field vortex solid- liquid transition is indeed the 
defect-unbinding transition, one ought to expect power-law $I(V)$ 
curves in the solid, $E 
\propto j^{a}$. Here, the power $a = (g/1+g) 2 \varepsilon_{0}(T)s /k_{B}T - 1$ 
is determined by the effective Coulomb gas ``charge'' $ [2 (g/1+g) 
\varepsilon_{0}s]^{1/2}$.\cite{Minnhagen95} Note that $g$ and therefore $a$ 
are, in principle, field--dependent; the exponential behavior of the 
irreversibility line suggests $a \propto 1 / \ln( B/B_{cr})$. Figure~\ref{fig:IV} shows that the 
$I(V)$--curves at intermediate temperature indeed follow a power--law. 
The power $\partial \log E / \partial \log j$ is plotted in fig.~\ref{fig:power}. A comparison
with the magnitude and temperature dependence expected for 
$\varepsilon_{0}(0)s/k_{B} = 770$ K found previously yields reasonable agreement 
if one sets $ g / 1 + g \approx 0.3$, {\em i.e.} $g \approx 0.4$. This value, although 
highly approximate ($g$ is expected to depend on field and 
temperature) suggests that intra-layer pancake vortex 
interactions and the lack of elastic relaxation of the pinned lattice 
is what determines both the dynamics and the vortex phase diagram.

\section{Conclusions}

We have produced large diameter (20 $\mu$m) short (200 nm)  damage 
tracks on a single surface of a \linebreak 
Bi$_{2}$Sr$_{2}$CaCu$_{2}$O$_{8+\delta}$ single crystal
by 30 MeV C$_{60}$ irradiation. After the irradiation, the vortex 
matter still shows the first order transition found in pristine 
crystal, but also the exponential irreversibility line normally found 
in heavy-ion irradiated Bi$_{2}$Sr$_{2}$CaCu$_{2}$O$_{8+\delta}$. 
Measurements of the flux distribution, magnetic relaxation, and 
$I(V)$--curves demonstrate exeptionally strong flux pinning. However, 
quantitative analysis of the data shows that neither the sustainable 
current density, nor the irreversibility line, depend on the pinning 
potential of the columnar tracks. More specifically , we suggest that 
the increase in current density after irradiation is due to the increase of the 
energy of a vacancy-interstitial pair in the pancake vortex latter in 
each layer.

\subsection*{Acknowledgments}
We are grateful to Y. Mehtar-Tani for the measurements on the pristine 
crystal.

\end{document}